\documentstyle[12pt]{article}

\begin{document}
\setcounter{page}{1} \pagestyle{plain} \vspace{1cm}
\begin{center}
\Large{\bf Implications of Minimal Length Scale on the Statistical
Mechanics of Ideal Gas }\\
\small
\vspace{1cm} {\bf Kourosh Nozari}\quad and \quad {\bf S. Hamid Mehdipour}\\
\vspace{0.5cm} {\it Department of Physics,
Faculty of Basic Science,\\
University of Mazandaran,\\
P. O. Box 47416-1467,
Babolsar, IRAN}\\
{\it e-mail: knozari@umz.ac.ir}

\end{center}
\vspace{1.5cm}

\begin{abstract}
Several alternative approaches to quantum gravity problem suggest
the modification of the {\it fundamental volume $\omega_{0}$} of the
accessible phase space for representative points. This modified
fundamental volume has a novel momentum dependence. In this paper,
we study the effects of this modification on the thermodynamics of
an ideal gas within the microcanonical ensemble and using the
generalized uncertainty principle(GUP). Although the induced
modifications are important only in quantum gravity era, possible
experimental manifestation of these effects may provides strong
support for underlying quantum gravity proposal.
\end{abstract}
\newpage

\section{Motivation}
In ordinary statistical mechanics, it is impossible to define the
position of a representative point in the phase space of the given
system more accurately than the situation which is given by $(\delta
q\, \delta p)_{min}\geq \hbar$. In another words, around any point
$(q,p)$ of the (two dimensional) phase space, there exists an area
of the order $\hbar$ which the position of the representative point
cannot be pin-pointed. In a phase space of $2N$ dimensions, the
corresponding volume of uncertainty around any point would be of
order $\hbar^{N}$. Therefore, it is reasonable to regard the phase
space as made up of elementary cells of volume $\approx\hbar^{N}$.
These cells have one-to-one correspondence with the quantum
mechanical states of the given system[1].\\
In ordinary picture of quantum theory, we usually ignore the
gravitational effects. Nevertheless, gravity induces uncertainty.
Combining this extra uncertainty with usual uncertainty principle of
Heisenberg, we find the generalized uncertainty principle(GUP).
Therefore, measurements in quantum gravity should be governed by
GUP. Much evidence exists (from string theory[2-6], non-commutative
geometry[7], loop quantum gravity[8] and black hole physics[9])
which confirm GUP. On the other hand, theories such as scale
relativity of Nottale[10], Twistors theory of Penrose[11] and
$E$-infinity of El Naschie[12-14] provide strong supports and deep
understanding of this finite resolution of spacetime points. All of
this evidence has its origin in the quantum fluctuations of the
background spacetime metric(the so-called spacetime
fuzziness[15,16]-spacetime fuzziness with given mathematical
exactness in fuzzy $K3$ manifold of ${\cal{E}^{\infty}}$[17]- and/or
foamy/fractal spacetime [18]). A common feature of all promising
candidates for quantum gravity is the existence of a minimal length
scale on the order of Planck length. This intriguing aspect of
quantum gravity has been investigated from different
perspectives[19-31].  A generalized uncertainty principle can be
formulated as
\begin{equation}
\delta x \delta p\geq{\hbar}\Big(1+\beta(\delta p)^{2}\Big).
\end{equation}
(Note that actually we should consider an extra term on the right
hand side of this equation which contains expectation value of $p$.
Since we are dealing with absolute value of non-vanishing minimal
uncertainty in position, we have set this term to be zero. For
complete discussion of this point see [15]). The main result of this
GUP is the existence of a non-vanishing minimal observable length
which is a consequence of finite resolution of spacetime at Planck
scale. In fact, position measurement is possible only up to a
multiple of Planck length, $l_{p}=\sqrt{\frac{G\hbar}{c^3}}\sim
10^{-33}cm$[15]. A simple calculation based on (1) shows that
$$(\Delta x)_{min}=\hbar\sqrt{\beta}.$$ Therefore $\beta$ is related
to minimal observable length. In the spirit of string theory this
minimal length is of the order of string length. It is impossible to
set up a measurement to find more accurate particle position than
Planck length, and this means that the notion of locality breaks
down(the so-called spacetime fuzziness)(see[32] and references
therein). For our purposes, this generalized uncertainty principle
can be rewritten as follows
\begin{equation}
\delta q \delta p\geq{\hbar}\Big(1+\beta(p)^{2}\Big).
\end{equation}
Note that one can interpret this result as a generalization of
$\hbar$ (this generalization can lead to generalization of De
Broglie principle[33]). Now it is obvious that the volume of
uncertainty around any point of phase space increases due to extra
term in the right hand side of (2)( or equivalently due to
generalization of $\hbar$). This generalized volume is given by
$\Big[{\hbar}\Big(1+\beta p^{2}\Big)\Big]^{N}$. Since $\beta$ is a
positive quantity, this feature will decrease drastically the number
of accessible microstates for a given system specially in high
momentum limit. Our goal here is to calculate this reduction within
microcanonical ensemble and investigate its consequences. We
consider a simple system of monatomic ideal gas in the case of
microcanonical ensemble. This issue firstly has been considered by
Kalyana Rama[34]. He has discussed the effect of GUP on various
thermodynamical quantities in {\it grand canonical ensemble}. Here
we are going to consider the effects of GUP on thermodynamics of
ideal gas in {\it microcanonical ensemble}. Note that an elegant
formulation of statistical mechanics of multi-dimensional Cantor
sets based on fractal nature of spacetime has been provided by El
Naschie[35].

In which follows we consider GUP as our primary input. Although GUP
is a model-independent concept, its functional form is quantum
gravity model dependent. As a result, there are more generalization
of GUP which consider further terms on the right hand side of
equation (1)(see [15]), but in some sense regarding dynamics,
equation (1) has more powerful physical grounds(from string
theoretical view point, see[6]). Note that GUP (1) has a
non-vanishing minimal uncertainty only in position; there is no
non-vanishing minimal uncertainty in momentum. Since aforementioned
reduction of accessible phase space is a quantum gravity effect, any
experimental test of quantum gravity signals will support our
findings.

\section{GUP and Thermodynamics of Ideal Gas}
To illustrate the concepts developed in the preceding section, we
derive here the thermodynamical properties of an ideal gas composed
of monatomic noninteracting particles within GUP. In microcanonical
ensemble, the macrostate of the given system is defined by the
number of molecules $N$, the volume $V$ and energy $E$ of the
system. In this ensemble, the volume $\omega$ of the phase space
accessible to the representative points of the (member) system where
have a choice to lie anywhere within a {\it hypershell} defined by
the condition $E-\frac{\Delta}{2}\leq H(q,p)\leq E+\frac{\Delta}{2}$
is given by
\begin{equation}
\omega=\int'd\omega=\int'\int'(d^{3N}q)\,(d^{3N}p)
\end{equation}
where $\omega\equiv\omega(N,V,E;\Delta)$, and the primed integration
extends only over that part of the phase space which conforms to the
above condition. Since the Hamiltonian in this case is a function of
the $p$'s alone, the integrations over the $q$'s can be carried out
straightforwardly which gives a factor of $V^N$. The remaining
integral
\begin{equation}
 \int^{\prime}(d^{3N}p)
\end{equation}
should be evaluated under the following condition
\begin{equation}
2m\bigg[E-\frac{\Delta}{2}\bigg]\leq\sum_{i=1}^{3N}p_{i}^2\leq
2m\bigg[E+\frac{\Delta}{2}\bigg].
\end{equation}
Now we should consider the following two key points[32,33],[36-38]:
\begin{itemize}
\item
Within GUP framework, particle's momentum generalizes. This
generalized momentum is given by
\begin{equation}
 p^{GUP}\simeq p\big(1+\frac{1}{3}\beta p^{2}\big)
\end{equation}
\item
Due to generalization of momentum, energy will generalize too
\begin{equation}
 E^{GUP}\simeq E\big(1+\frac{1}{3}\beta E^{2}\big).
\end{equation}
\end{itemize}
Note that for simplicity we have considered only first order
corrections. Higher order corrections lead to integrals which can be
calculated only with sophisticated numerical scheme. With this two
points in mind, up to first order in $\beta$, the {\it hypershell}
equation is given by
\begin{equation}
2m\bigg[E\big(1+\frac{1}{3}\beta
E^{2}\big)-\frac{\Delta}{2}\bigg]\leq\sum_{i=1}^{3N}p_i^2\big(1+\frac{2}{3}\beta
p_{i}^{2}\big)\leq2m\bigg[E\big(1+\frac{1}{3}\beta
E^{2}\big)+\frac{\Delta}{2}\bigg].
\end{equation}
Now, integral (4) is equal to the volume of a $3N$-dimensional
hypershell, bounded by two hyperspheres of radii
$$\sqrt{2m\bigg[E\big(1+\frac{1}{3}\beta
E^{2}\big)-\frac{\Delta}{2}\bigg]} \qquad and \qquad
\sqrt{2m\bigg[E\big(1+\frac{1}{3}\beta
E^{2}\big)+\frac{\Delta}{2}\bigg]} .$$ Therefore, we can write
\begin{equation}
\int'\ldots\int'\prod_{i=1}^{3N}dp_i=C_{3N}\Bigg(\sqrt{2m(E^{GUP}+\frac{\Delta}{2})}\Bigg)^{3N}:=\Lambda
\end{equation}
where
$$0\leq\sum_{i=1}^{3N}p_i^2(1+\frac{2}{3}\beta
p_i^2)\leq 2m(E^{GUP}+\frac{\Delta}{2}).$$ This statement gives half
of the volume of relevant phase space(since $\Delta\ll E^{GUP}$) and
therefore we should multiply our final result with a factor of $2$.
Here $C_{3N}$ is a constant which depends only on the dimensionality
of phase space. Clearly, the volume element $d\Lambda$ can also be
written as
\begin{equation}
d\Lambda=\frac{3}{2}NC_{3N}\Big(\sqrt{2m}\Big)^{3N}
\Bigg[\sqrt{E^{GUP}+\frac{\Delta}{2}}\Bigg]^{3N-2}dE^{GUP}.
\end{equation}
To evaluate $C_{3N}$, we use the following integral formula,
\begin{equation}
\int_{-\infty}^{+\infty} \exp(-p^2-\frac{2}{3}\beta
p^4)dp=\frac{1}{2}\Big(\frac{3}{2\beta}\Big)^{\frac{1}{2}}\exp\Big(\frac{3}{16\beta}\Big)
K_{\frac{1}{4}}\Big(\frac{3}{16\beta}\Big),
\end{equation}
where $K_\nu(x)$ is the modified Bessel function of the second kind.
Multiplying $3N$ such integrals, one for each of variables $p_i$, we
obtain
\begin{equation}
\int_{-\infty}^{+\infty}\ldots\int_{-\infty}^\infty
\exp\Big(-\sum_{i=1}^{3N}p_i^2(1+\frac{2}{3}\beta
p_i^2)\Big)\prod_{i=1}^{3N}dp_i=\bigg[\frac{1}{2}\Big(\frac{3}{2\beta}\Big)^{\frac{1}{2}}\exp\Big(\frac{3}{16\beta}\Big)
K_{\frac{1}{4}}\Big(\frac{3}{16\beta}\Big)\bigg]^{3N}.
\end{equation}
Therefore, it follows that
\begin{equation}
\int_{-\infty}^{+\infty}
\exp\bigg(-2m\Big(E^{GUP}+\frac{\Delta}{2}\Big)\bigg)d\Lambda=\bigg[\frac{1}{2}\Big(\frac{3}{2\beta}\Big)^
{\frac{1}{2}}\exp\Big(\frac{3}{16\beta}\Big)
K_{\frac{1}{4}}\Big(\frac{3}{16\beta}\Big)\bigg]^{3N}.
\end{equation}
Using equation (10), we find
\begin{equation}
\int_{-\infty}^{+\infty}
\frac{3}{2}NC_{3N}\Big(\sqrt{2m}\Big)^{3N}\Bigg[\sqrt{E^{GUP}+\frac{\Delta}{2}}\Bigg]^{3N-2}
\exp\bigg(-2m\Big(E^{GUP}+\frac{\Delta}{2}\Big)\bigg)dE^{GUP}=[n(\beta)]^{3N},
\end{equation}
where for simplicity we have defined
$n(\beta)=\frac{1}{2}\Big(\frac{3}{2\beta}\Big)^{\frac{1}{2}}\exp\Big(\frac{3}{16\beta}\Big)
K_{\frac{1}{4}}\Big(\frac{3}{16\beta}\Big)$. Calculation of this
integral gives
\begin{equation}
C_{3N}=\frac{2[n(\beta)]^{3N}\exp(m\Delta)}{3N(2m)^{\frac{3N}{2}}\int_{-\infty}^{+\infty}
(E^{GUP}+\frac{\Delta}{2})^{\frac{3N-2}{2}}\exp(-2mE^{GUP})dE^{GUP}}.
\end{equation}
For $\Delta\ll E^{GUP}$, this equation reduces to
\begin{equation}
C_{3N}=\frac{2[n(\beta)]^{3N}}{3N(\frac{3N}{2}-1)!}.
\end{equation}
Now, from equation (9), we find (since $N\gg 1$)
\begin{equation}
\int\ldots\int\prod_{i=1}^{3N}dp_i\equiv\frac{2[n(\beta)]^{3N}(2mE^{GUP})^{\frac{3N}{2}}
\Big[1+\frac{3N\Delta}{4E^{GUP}}\Big]}{3N(\frac{3N}{2}-1)!}\simeq\frac{\Delta}{2E^{GUP}}
\frac{[n(\beta)]^{3N}}{(\frac{3N}{2}-1)!}
\Big(2mE^{GUP}\Big)^{\frac{3N}{2}}.
\end{equation}
Hence, the total volume of the phase space enclosed within
hypershell is given by
\begin{equation}
\omega\simeq\frac{\Delta}{E^{GUP}}V^N
\frac{\Big(2[n(\beta)]^2mE^{GUP}\Big)^{\frac{3N}{2}}}{(\frac{3N}{2}-1)!}.
\end{equation}
There exists a direct correspondence between the various microstates
of the given system and the various locations in the phase space.
The volume $\omega$ (of the allowed region of the phase space) is,
therefore, a direct measure of the multiplicity $\Omega$ of the
microstates obtaining in the system. To establish a numerical
correspondence between $\omega$ and $\Omega$, we should discover a
{\it fundamental volume} $\omega_0$ which could be regarded as
equivalent to one microstate. Once this is done, we can right away
conclude that, asymptotically,
\begin{equation}
\Omega=\frac{\omega}{\omega_0},
\end{equation}
where in GUP, $\omega_0$ is given by
\begin{equation}
\omega_0=(\delta q\,\delta p)^{3N}=[\hbar(1+\beta p^2)]^{3N}\equiv
\hbar'^{3N},
\end{equation}
Finally, we find the multiplicity(total number of microstates) of
the given system as
\begin{equation}
\Omega=\frac{V^N}{\hbar'^{3N}}\,\frac{\Delta}{E^{GUP}}\,
\frac{\Big(2[n(\beta)]^2mE^{GUP}\Big)^{\frac{3N}{2}}}{(\frac{3N}{2}-1)!}.
\end{equation}
Apparently, within GUP, due to increased fundamental volume
$\omega_{0}$, number of total microstates decreases. This point have
been discussed in more powerful manner in terms of information loss
on the quantum scale of spacetime by El Naschie[35]. In fact based
on El Naschie point of view, there is a fuzzy region information
related to the de Broglie length followed by an informational
horizon given by the loss of information. In this manner he find a
bound for this information loss(see reference [35] page 207).

The complete thermodynamics of the given system would then follow in
the usual way, namely through the relationship,
\begin{equation}
S(N,V,E^{GUP})=k\ln\Omega=k\ln\Bigg(\frac{V^N}{\hbar'^{3N}}\,\frac{\Delta}{E^{GUP}}\,
\frac{\Big(2[n(\beta)]^2mE^{GUP}\Big)^{\frac{3N}{2}}}{(\frac{3N}{2}-1)!}\Bigg).
\end{equation}
In the absence of quantum gravitational effects, that is when
$\beta=0$, we obtain standard thermodynamical results(note that
$n(\beta)$ tends to $\sqrt{\pi}$ when $\beta\rightarrow 0$). Various
thermodynamical quantities can then be calculated using equation(22)
straightforwardly.\\
Our analysis shows a reduction of the number of accessible
microstates in high momentum regime. This reduction of has novel
implications such as reduction of corresponding microcanonical
entropy of the system. It seems that thermodynamical systems at very
short distances have an "unusual thermodynamics". Recently, we have
shown[39], for a micro-black hole which lies in the spirit of such
very small scale systems, that the temperature has an unusual
behavior and entropy tends to zero when the size of the system tends
to Planck length. Here, it is evident that in high momentum regime
the entropy of ideal gas tends to zero which resembles the mentioned
behavior of micro-black hole. This seems to be a universal behavior
of short distance physics.\\
Note that our analysis is based on the generalization of momentum
$p$ and energy $E$(see equations (6) and (7)). These generalizations
lead to modified dispersion relations. Modification of dispersion
relations is motivated by several quantum gravity scenarios(see [32]
and [36-38]). Searches for modifications of the dispersion relation
of the form (6) or (7) (and their more general forms), constitute
part of experimental quantum gravity efforts. This type of effect is
beyond the standard model extension, since it could not be described
by power counting renormalizable terms. A possible manifestation of
the modified dispersion relations is an energy dependent propagation
velocity which should lead to different time-of-arrivals of the same
event on a distant star when looked at it in different frequency
channels[40](also see [32] and references therein). Another effect
is frequency dependent position of interference fringes in
interferometric experiments. These are some possible experimental
schemes for detecting such quantum gravity signals. Although these
experimental scheme are not direct consequence of our findings, but
their possible support of modified dispersion relations can be
considered as an indirect support of our finding. Therefore it is
possible in principle to test reduction of phase space due to
quantum gravity effects.

\section{GUP and Thermodynamics of Extreme Relativistic Gas}
In this section, we calculate thermodynamics of an
ultra-relativistic monatomic noninteracting gaseous system within
GUP and up to first order corrections. Using arguments presented in
preceding section, the hypershell equation for ultra-relativistic
gaseous system is given by
\begin{equation}
\frac{1}{c}\bigg[E\big(1+\frac{1}{3}\beta
E^{2}\big)-\frac{\Delta}{2}\bigg]\leq\sum_{i=1}^{3N}p_i\big(1+\frac{1}{3}\beta
p_{i}^{2}\big)\leq\frac{1}{c}\bigg[E\big(1+\frac{1}{3}\beta
E^{2}\big)+\frac{\Delta}{2}\bigg].
\end{equation}
Now integral (4) is equal to the volume of a $3N$-dimensional
hypershell, bounded by two hyperspheres of radii
$$\sqrt{\frac{1}{c}\bigg[E\big(1+\frac{1}{3}\beta
E^{2}\big)-\frac{\Delta}{2}\bigg]} \qquad and \qquad
\sqrt{\frac{1}{c}\bigg[E\big(1+\frac{1}{3}\beta
E^{2}\big)+\frac{\Delta}{2}\bigg]}.$$\\
The number of accessible microstates for the system is proportional
to the volume of this hypershell.  In the same manner as previous
section, we have
\begin{equation}
\int'\ldots\int'\prod_{i=1}^{3N}dp_i=D_{3N}\Bigg(\sqrt{\frac{1}{c}(E^{GUP}+\frac{\Delta}{2})}\Bigg)^{3N}
:=\Upsilon,
\end{equation}
where $$0\leq\sum_{i=1}^{3N}p_i(1+\frac{1}{3}\beta
p_i^2)\leq\frac{1}{c}(E^{GUP}+\frac{\Delta}{2}).$$ Here $D_{3N}$, is
a constant which depends only to the dimensionality of the phase
space. Now, since $d\Upsilon$ can be written as
\begin{equation}
d\Upsilon=\frac{3}{2}ND_{3N}\Big(\frac{1}{c}\Big)^{\frac{3N}{2}}\Bigg[\sqrt{E^{GUP}+\frac{\Delta}{2}}
\Bigg]^{3N-2}dE^{GUP},
\end{equation}
to evaluate $D_{3N}$, we use the following integral formula
\begin{equation}
\int_{-\infty}^{+\infty} \exp(-p-\frac{1}{3}\beta p^3)dp=m(\beta),
\end{equation}
where,
$m(\beta)=\frac{2}{3\sqrt{\beta}}\Big[{\cal{L}}S2(0,\frac{1}{3},\frac{2}{3\sqrt{\beta}})
+{\cal{L}}S2(0,\frac{1}{3},-\frac{2}{3\sqrt{\beta}})\Big]$  is a
linear combination of Lommel functions. Note that we should consider
the real part of this combination. Multiplying $3N$ such integrals,
one for each of variables $p_i$, we obtain
\begin{equation}
\int_{-\infty}^{+\infty}\ldots\int_{-\infty}^\infty
\exp\Big(-\sum_{i=1}^{3N}p_i(1+\frac{1}{3}\beta
p_i^2)\Big)\prod_{i=1}^{3N}dp_i=[m(\beta)]^{3N},
\end{equation}
whence it follows that,
\begin{equation}
\int_{-\infty}^{+\infty}
\exp\bigg(-\frac{1}{c}\Big(E^{GUP}+\frac{\Delta}{2}\Big)\bigg)dR_{3N}(R)=[m(\beta)]^{3N}.
\end{equation}
Now, using equation(25), we find
\begin{equation}
\int_{-\infty}^{+\infty}
\frac{3}{2}ND_{3N}\Big(\frac{1}{c}\Big)^{\frac{3N}{2}}\Bigg[\sqrt{E^{GUP}+\frac{\Delta}{2}}\Bigg]^{3N-2}\exp\bigg
(-\frac{1}{c}\Big(E^{GUP}+\frac{\Delta}{2}\Big)\bigg)dE^{GUP}
=[m(\beta)]^{3N}.
\end{equation}
Therefore $D_{3N}$ is given by
\begin{equation}
D_{3N}=\frac{2[m(\beta)]^{3N}\exp\Big(\frac{\Delta}{2c}\Big)}{3N(\frac{1}{c})^{\frac{3N}{2}}
\int_{-\infty}^{+\infty}(E^{GUP}+\frac{\Delta}{2})^{\frac{3N-2}{2}}\exp(-\frac{1}{c}E^{GUP})dE^{GUP}}.
\end{equation}
For $\Delta\ll E^{GUP}$ one obtains
\begin{equation}
D_{3N}=\frac{2[m(\beta)]^{3N}}{3N(\frac{3N}{2}-1)!}.
\end{equation}
Now from equation (24) we have
\begin{equation}
\int\ldots\int\prod_{i=1}^{3N}dp_i\equiv\frac{2[m(\beta)]^{3N}(\frac{1}{c}E^{GUP})^{\frac{3N}{2}}
\Big[1+\frac{3N\Delta}{4E^{GUP}}\Big]}{3N(\frac{3N}{2}-1)!}\simeq\frac{\Delta}{2E^{GUP}}\frac{[m(\beta)]^{3N}}
{(\frac{3N}{2}-1)!}\bigg(\frac{1}{c}E^{GUP}\bigg)^{\frac{3N}{2}}.
\end{equation}
Hence, the total volume of the phase space enclosed within
hypershell is given by
\begin{equation}
\omega\simeq\frac{\Delta}{E^{GUP}}V^N
\frac{\Big(\frac{1}{c}[m(\beta)]^2E^{GUP}\Big)^{\frac{3N}{2}}}{(\frac{3N}{2}-1)!}.
\end{equation}
Since
\begin{equation}
\Omega=\frac{\omega}{\omega_0},
\end{equation}
we find finally
\begin{equation}
\Omega=\frac{V^N}{\hbar'^{3N}}\,\frac{\Delta}{E^{GUP}}\,
\frac{\Big(\frac{1}{c}[m(\beta)]^2E^{GUP}\Big)^{\frac{3N}{2}}}{(\frac{3N}{2}-1)!}.
\end{equation}
The thermodynamics of the system would then follow in the usual way
through the following relation,
\begin{equation}
S(N,V,E^{GUP})=k\ln\Omega=k\ln\Bigg(\frac{V^N}{\hbar'^{3N}}\,\frac{\Delta}{E^{GUP}}\,
\frac{\Big(\frac{1}{c}[m(\beta)]^2E^{GUP}\Big)^{\frac{3N}{2}}}{(\frac{3N}{2}-1)!}\Bigg).
\end{equation}
In the standard situation where $\beta=0$, we obtain well-known
results of ordinary statistical mechanics. Various thermodynamical
quantities can then be calculated using equation(36). Once again,
number of total microstates decreases leading to less entropy for
given system in GUP.

\section{Summary and Discussion}
Generalized uncertainty principle(GUP) and/or modified dispersion
relations(MDRs) are common features of all promising quantum gravity
scenarios. As a result of GUP and/or MDRs, the fundamental volume
$\omega_{0}$ of accessible phase space for representative points of
a given statistical system increases due to gravitational
uncertainty. This increasing fundamental volume of accessible phase
space can be interpreted as a result of Planck constant
generalization. This feature requires a reformulation of statistical
mechanics since number of accessible microstates for given system
decreases drastically in quantum gravity era. This reduction of
microstates is quantum gravity effect and can be avoided in standard
situation but it plays important role in thermodynamics of early
universe. Here we have considered an ideal gaseous system composed
of noninteracting monatomic molecules within microcanonical
ensemble. We have shown that GUP as a manifestation of quantum
nature of gravity, leads to reduction of accessible microstates of
the given system. We have considered two limits of classical and
ultra-relativistic gas and in each case we have computed complete
thermodynamics of the system.\\
Our analysis shows a reduction of the number of accessible
microstates in high momentum regime. This reduction of has novel
implications such as reduction of corresponding microcanonical
entropy of the system. In the limit of very high momentum, the
entropy of the system tends to zero. It seems that thermodynamical
systems at very short distances(very high energy) have an "unusual
thermodynamics".\\
One can ask about the possible detection of these extra-ordinary
effects. Up to now, there is no direct experimental or observational
scheme for detection of these novel effects. Nevertheless, since the
basis of our calculations come back to GUP and/or MDRs, possible
experimental schemes for detecting these quantum gravity signals are
indirect tests of our findings. Several strategies for testing these
quantum gravity predictions(GUP and/or MDRs) have been proposed[38].
Strategies such as frequency dependent position of interference
fringes in interferometry experiments, an energy dependent
propagation velocity which should lead to different time-of-arrivals
of the same event on a distant star when looked at it in different
frequency channels, possible detection of black hole remnants in
Large Hadronic Collider (LHC) and also in ultrahigh energy cosmic
ray (UHECR) air showers[41,42], are possible evidences for testing
these novel quantum gravity effects. Therefore, any search for
quantum gravity signals provides possible indirect test of
generalized statistical mechanics which we have constructed.


\begin{thebibliography}{10}
\bibitem{1}
R. K.Pathria, {\it Statistical Mechanics}, Pergamon Press, First
Edition, 1972
\bibitem{2}
G. Veneziano,  A stringy nature needs just two constants, {\it
Europhys. Lett.}{\bf 2} (1986) 199; Proc. of Texas Superstring
Workshop (1989).
\bibitem{3}
D. Amati, M. Ciafaloni and G. Veneziano,  Can spacetime be probed
below the string size?, {\it Phys. Lett. B} {\bf 216} (1989) 41
\bibitem{4}
D. Amati, M. Ciafaloni and G. Veneziano, Superstring collisions at
Planckian energies, {\it Phys. Lett. B}{\bf197} (1987) 81; Classical
and quantum gravity effects from Planckian energy superstring
collisions,  {\it Int. J. Mod. Phys. A}{\bf 7} (1988) 1615;
Higher-order gravitational deflection and soft bremsstrahlung in
Planckian energy superstring collisions,  {\it Nucl. Phys.
B}{\bf347} (1990) 530.
\bibitem{5}
D. J. Gross and P. F. Mende,  String theory beyond the Planck scale,
{\it Nucl. Phys. B} {\bf 303} (1988)407
\bibitem{6}
K. Konishi, G. Paffuti and P. Provero,  Minimum physical length and
the generalized uncertainty principle in string theory,  {\it Phys.
Lett. B} {\bf 234} (1990) 276
\bibitem{7}
S. Capozziello, G. Lambiase, G. Scarpetta,  Generalized Uncertainty
Principle from Quantum Geometry,  {\it Int. J. Theor. Phys.}{\bf 39}
(2000) 15.
\bibitem{8}
L. J. Garay, Quantum gravity and Minimum length,  {\it Int. J. Mod.
Phys. A}{\bf10} (1995) 145
\bibitem{9}
F. Scardigli,  Generalized Uncertainty Principle in Quantum Gravity
from Micro-Black Hole Gedanken Experiment,  {\it Phys. Lett.
B}{\bf452} (1999) 39-44
\bibitem{10}
L. Nottale, {\it Fractal space-time and microphysics: towards a
theory of scale relativity}, World Scientific, Singapore, 1993
\bibitem{11}
R. Penrose,  The Central Programme of Twistor Theory, {\it Chaos,
Solitons and Fractals}, {\bf 10} (1999) 581-611
\bibitem{12}
M. S. El Naschie, The concepts of E-infinity: An elementary
introduction to the Cantorian-fractal theory of quantum physics,
{\it Chaos, Solitons and Fractals} {\bf 22} (2004) 495-511
\bibitem{13}
M. S. El Naschie,  Elementary prerequisites for
E-infinity(Recommended background readings in nonlinear dynamics,
geometry and topology), {\it Chaos, Solitons and Fractals}, {\bf 30}
(2006) 579-605
\bibitem{14}
- M. S. El Naschie, Intermediate prerequisites for
E-infinity(Further recommended reading in nonlinear dynamics and
mathematical physics), {\it Chaos, Solitons and Fractals}, {\bf 30}
(2006) 622-628\\
- M. S. El Naschie, Advanced prerequisites for E-infinity theory,
{\it Chaos, Solitons and Fractals}, {\bf 30} (2006) 636-641
\bibitem{15}
A. Kempf, {\it et al.},  Hilbert space representation of the minimal
length uncertainty relation, {\it Phys. Rev. D}{\bf52} (1995) 1108.
\bibitem{16}
K. Nozari and B. Fazlpour, Some consequences of spacetime fuzziness,
{\it Chaos, Solitons and Fractals},(2006) (in Press)
\bibitem{17}
M. S. El Naschie, Fuzzy Dodecahedron topology and E-infinity
spacetime as a model for quantum physics, {\it Chaos, Solitons and
Fractals}, {\bf 30} (2006) 1025-1033;  Topics in the mathematical
physics of E-infinity theory, {\it Chaos, Solitons and Fractals},
{\bf 30} (2006) 656-663;  On two new fuzzy K\"{a}hler manifolds,
Klein modular space and t' Hooft holographic principles, {\it Chaos,
Solitons and Fractals}, {\bf 29} (2006) 876-881
\bibitem{18}
G. Amelino-Camelia,  A phenomenological description of
quantum-gravity-induced space-time noise,  {\it Nature} {\bf 410}
(2001) 1065-1069
\bibitem{19}
M. Maggiore,  A Generalized Uncertainty Principle in Quantum Gravity
, {\it Phys. Lett. B}{\bf 304} (1993) 65.
\bibitem{20}
M. Maggiore,  Quantum Groups, Gravity, and the Generalized
Uncertainty Principle,  {\it Phys. Rev. D}{\bf49} (1994) 5182.
\bibitem{21}
F. Scardigli, R. Casadio,  Generalized Uncertainty Principle,
Extra-dimensions and Holography,  {\it Class. Quant. Grav.}{\bf 20}
(2003) 3915
\bibitem{22}
S. Hossenfelder,  Running Coupling with Minimal Length, {\it Phys.
Rev. D}{\bf70} (2004) 105003; Interpretation of Quantum Field
Theories with a Minimal Length Scale, {\it Phys. Rev. D} {\bf73}
(2006) 105013
\bibitem{23}
S. Hossenfelder {\it et al},  Signatures in the Planck Regime, {\it
Phys. Lett. B}{\bf 575} (2003) 85-99
\bibitem{24}
R. Akhoury and Y. P. Yao, Minimal Length Uncertainty Relation and
the Hydrogen Spectrum, {\it Phys. Lett. B}{\bf572} (2003) 37-42
\bibitem{25}
K. Nozari and T. Azizi, Some Aspects of Minimal Length Quantum
Mechanics, {\it Gen. Rel. Grav.} {\bf 38} (2006) 735-742
\bibitem{26}
K. Nozari and M. Karami, Minimal Length and Generalized Dirac
Equation,  {\it Mod. Phys. Lett. A} {\bf20}(2005) 3095-3104
\bibitem{27}
K. Nozari and S. H. Mehdipour, Gravitational Uncertainty and Black
Holes Remnants, {\it Mod. Phys. Lett. A} {\bf 20} (2005) 2937-2948
\bibitem{28}
Kourosh Nozari, Some Aspects of Planck Scale Quantum Optics, {\it
Phys. Lett. B} {\bf 629} (2005) 41-52
\bibitem{29}
M. Lubo, Quantum Minimal Length and Transplanckian Photons, {\it
Phys. Rev. D}{\bf61} (2000) 124009; Mimimal Length Uncertainty
Principle and the Transplanckian Problem of Black Hole Physics, {\it
Phys. Rev. D} {\bf59} (1999) 044005
\bibitem{30}
S. Benczik {\it et al}, Short Distance vs. Long Distance Physics:
The Classical Limit of the Minimal Length Uncertainty Relation, {\it
Phys. Rev. D} {\bf66} (2002) 026003
\bibitem{31}
L. N. Chang {\it et al}, The Effect of the Minimal Length
Uncertainty Relation on the Density of States and the Cosmological
Constant Problem, {\it Phys. Rev. D}{\bf65} (2002) 125028
\bibitem{32}
G. Amelino-Camelia {\it et al}, The Search for Quantum Gravity
Signals,  arXiv:gr-qc/0501053 (AIP Conference Proceedings of the 2nd
Mexican Meeting on Mathematical and Experimental Physics).
\bibitem{33}
K. Nozari and S. H. Mehdipour, Wave Packets Propagation in Quantum
Gravity, {\it Gen. Rel. Grav.}{\bf 37} (2005) 1995
\bibitem{34}
S. Kalyana Rama, Some Consequences of the Generalised Uncertainty
Principle: Statistical Mechanical, Cosmological, and Varying Speed
of Light, {\it Phys. Lett. B} {\bf 519} (2001) 103-110.
\bibitem{35}
M. S. El Naschie, Statistical Mechanics of Multi-Dimensional Cantor
Sets, G\"{o}del Theorem and Quantum Spacetime, {\it Journal of the
Franklin Institute} {\bf 330} (1993) 199-211
\bibitem{36}
R. Gambini and J. Pullin, {\it Phys. Rev. D} {\bf 59} (1999)124021
\bibitem{37}
J. Alfaro {\it et al}, {\it Phys. Rev. Lett.} {\bf 84} (2000) 2318
\bibitem{38}
G. Amelino-Camelia {\it et al}, {\it Phys. Rev. D} {\bf 70} (2004)
107501
\bibitem{39}
K. Nozari and B. Fazlpour, Thermodynamics of an Evaporating
Schwarzschild Black Hole in Noncommutative Space,
arxiv:hep-th/0605109
\bibitem{40}
B. E. Schaefer, {\it Phys. Rev. Lett.} {\bf 82} (1999) 4964
\bibitem{41}
http://lhc-new-homepage.web.cern.ch/lhc-new-homepage/
\bibitem{42}
J. L. Feng and A. D. Shapere, {\it Phys. Rev. Lett.}{\bf 88} (2002)
021303; A. Ringwald and H. Tu, {\it Phys. Lett. B}{\bf
525}(2002)135; {\it Phys. Lett. B}{\bf 529} (2002) 1; M. Ave, E. J.
Ahn, M. Cavagli`a and A. V. Olinto, {\it Phys. Rev. D}{\bf
68}(2003)043004; S. I. Dutta, M. H. Reno and I. Sarcevic, {\it Phys.
Rev. D}{\bf 66} (2002)033002; R. Emparan, M. Masip and R. Rattazzi,
{\it Phys. Rev. D}{\bf 65}(2002)064023; A. Mironov, A. Morozov and
T. N. Tomaras, arXiv:hep-ph/0311318; P. Jain, S. Kar, S. Panda and
J. P. Ralston, {\it Int. J. Mod. Phys. D}{\bf 12}(2003)1593; M.
Cavagli`a, {\it Int. J. Mod. Phys. A}{\bf18} (2003) 1843; G.
Landsberg, arXiv:hep-ex/0310034; P. Kanti, arXiv:hep-ph/0402168.
\end{thebibliography}
\end{document}